\documentclass[a4paper,fleqn,usenatbib]{mnras}

\usepackage{newtxtext,newtxmath}
\usepackage[T1]{fontenc}
\usepackage{ae,aecompl}

\usepackage{graphicx}	% Including figure files
\usepackage{amsmath}	% Advanced maths commands
\usepackage{amssymb}	% Extra maths symbols

\title[Phosphorus-bearing molecules in the Galactic Center]{Phosphorus-bearing molecules in the Galactic Center}

\author[Rivilla et al.]{
V. M. Rivilla$^{1}$, %\thanks{E-mail: rivilla@arcetri.astro.it},
I. Jim\'enez-Serra$^{2}$,
S. Zeng$^{2}$,
S. Mart\'in$^{3,4}$,
J. Mart\'in-Pintado$^{5}$,
\newauthor
J.  Armijos-Abenda\~no$^{6}$,
S. Viti$^{7}$,
R. Aladro$^{8}$,
D. Riquelme$^{8}$,
\newauthor 
M. Requena-Torres$^{9}$,
D. Qu\'enard$^{2}$,
F. Fontani$^{1}$,
and M. T. Beltr\'an$^{1}$
\\
$^{1}$INAF/Osservatorio Astrofisico di Arcetri, Largo Enrico Fermi 5, I-50125, Florence, Italy\\
$^{2}$School of Physics and Astronomy, Queen Mary University of London, Mile End Road, London E1 4NS\\
$^{3}$Joint ALMA Observatory, Alonso de C\'ordova 3107, Vitacura 763 0355, Santiago, Chile \\
$^{4}$European Southern Observatory, Alonso de C\'ordova 3107, Vitacura Casilla 763 0355, Santiago, Chile\\
$^{5}$Centro de Astrobiolog\'ia (INTA-CSIC), Ctra. de Ajalvir Km. 4, Torrej\'on de Ardoz, 28850 Madrid, Spain\\
$^{6}$Observatorio Astron\'omico de Quito, Escuela Polit\'ecnica Nacional, Av. Gran Colombia S/N y Av. Diez de Agosto, Quito 170403, Ecuador\\
$^{7}$Department of Physics and Astronomy, UCL, Gower St., London, WC1E 6BT, UK\\
$^{8}$Max-Planck-Institut f\"ur Radioastronomie, Auf dem H\"ugel 69, 53121 Bonn, Germany \\
$^{9}$Space Telescope Science Institute, 3700 San Martin Drive, Baltimore, MD21218, USA\\
}

\date{Accepted XXX. Received YYY; in original form ZZZ}

\pubyear{2017}

\begin{document}
\label{firstpage}
\pagerange{\pageref{firstpage}--\pageref{lastpage}}
\maketitle

\begin{abstract}
Phosphorus (P) is one of the essential elements for life due to its central role in biochemical processes.  
Recent searches have shown that P-bearing molecules (in particular PN and PO) are present in star-forming regions, although their formation routes remain poorly understood. In this Letter, we report observations of PN and PO towards seven molecular clouds located in the Galactic Center, which are characterized by different types of chemistry. PN is detected in five out of seven sources, whose chemistry is thought to be shock-dominated. The two sources with PN non-detections correspond to clouds exposed to intense UV/X-rays/cosmic-ray radiation. 
PO is detected only towards the cloud G+0.693$-$0.03, with a PO/PN abundance ratio of $\sim$1.5.
%PN abundances show a clear correlation with the abundances of the optically-thin shock tracer $^{29}$SiO, which supports the idea that PN is efficiently enhanced in shocks.   
We conclude that P-bearing molecules likely form in shocked gas as a result of dust grain sputtering, while are destroyed by intense UV/X-ray/cosmic ray radiation.
%Until recently, P-bearing molecules in space had mostly been detected towards the envelopes of evolved stars.
%which is the richest source of O-bearing molecules in the Galactic Center. 
%Our chemical modeling of PN and PO in clouds affected by shocks and in UV-photon/cosmic-ray/X-ray illuminated clouds, confirms the observed trend for PN.
\end{abstract}


%%%%%%%%%%%%%%%%%%%%%%%%%%%%%%%%%%%%%%%%%%%%%%%%%%

%%%%%%%%%%%%%%%%% BODY OF PAPER %%%%%%%%%%%%%%%%%%

\section{Introduction}

\begin{figure*}
\includegraphics[height=5cm, width=10.2cm]{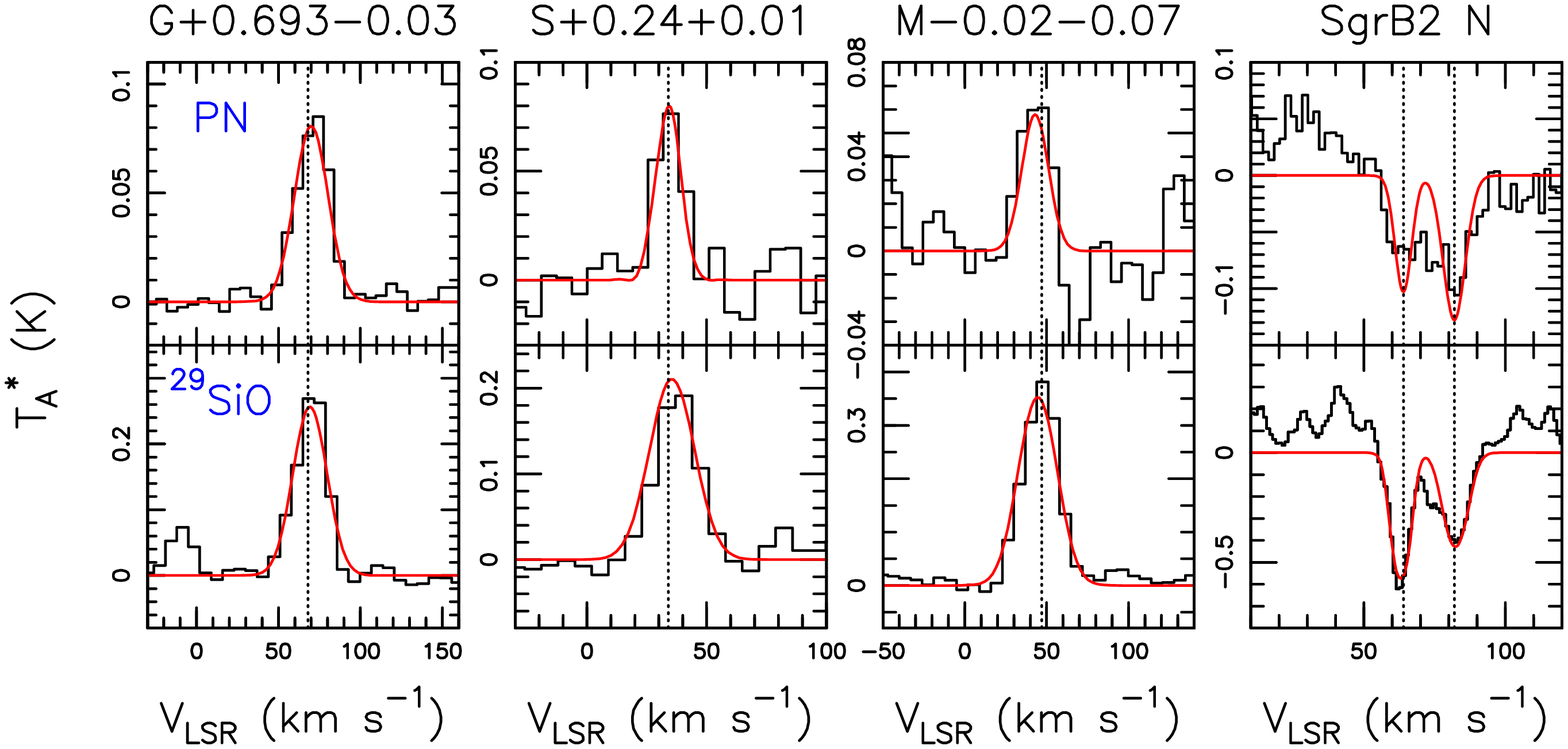}
\hspace{0.035cm}
\includegraphics[height=5.1cm, width=7.2cm]{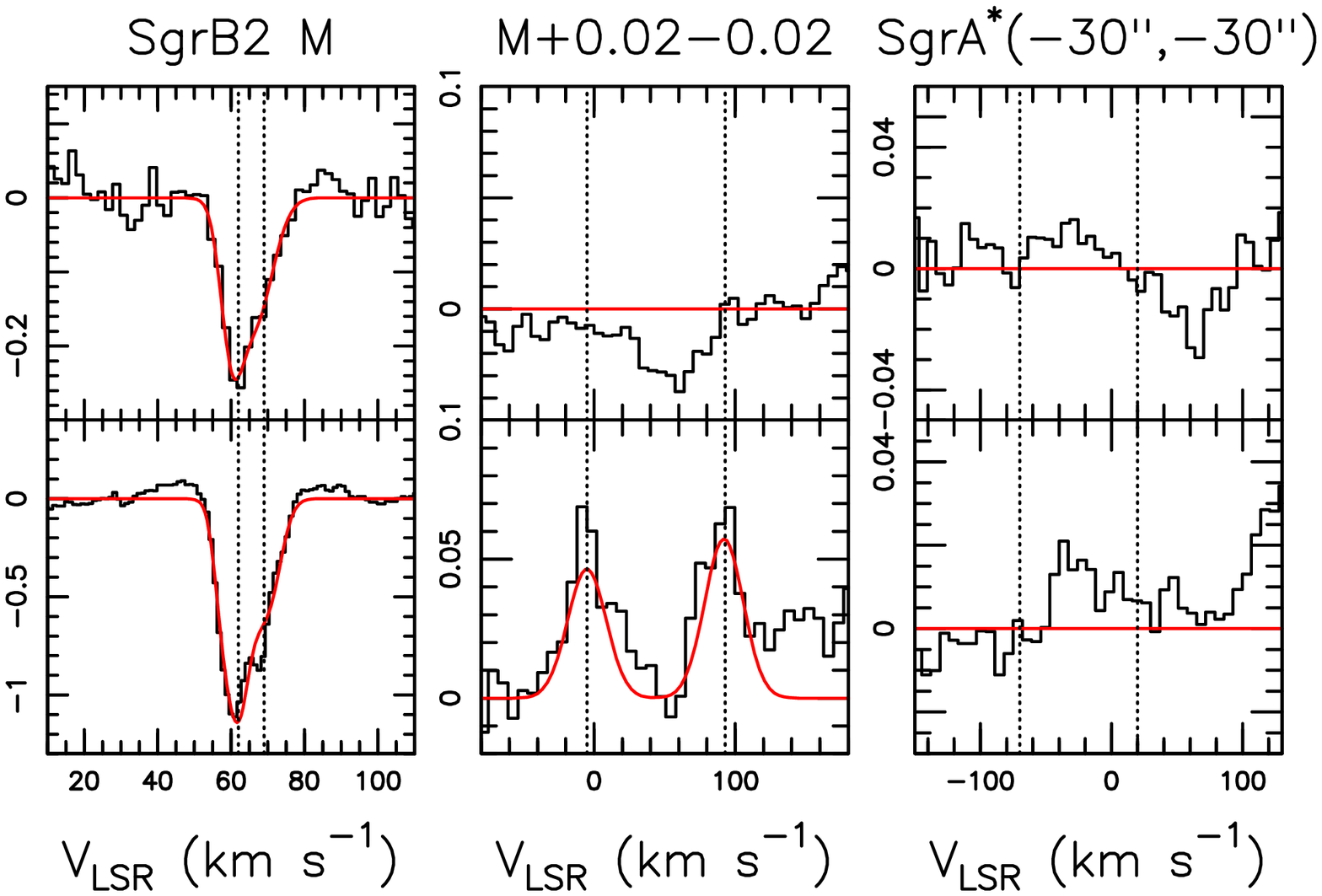}
 \caption{PN (2$-$1) and  $^{29}$SiO (2$-$1) lines measured towards all sources. The LTE best fits from MADCUBA are also shown (red lines).}
    \label{fig-line-profiles}
\end{figure*}

Phosphorus (P) is essential for life because it plays a central role in the formation of macromolecules such as phospholipids (the structural components of cellular membranes) and the deoxyribonucleic acid 
\citep[DNA,][]{macia1997}. It is synthesized in massive stars \citep{koo2013}, and it has relatively low cosmic abundance relative to hydrogen (2.8$\times$10$^{-7}$; \citealt{grevesse1998}). It is believed to be heavily depleted in cold and dense molecular clouds \citep{turner1990,wakelam2008}. 
%(by factors 600$-$10$^{4}$) since it is barely detected in the ISM \citep{turner1990,wakelam2008}. 
%The ion P$^+$ has only been detected in diffuse clouds \citep{jura1978}, and, 
Until recently, only a few simple P-bearing species (PN, PO, CP, HCP, C$_2$P, PH$_3$) had been identified towards the envelopes of evolved stars \citep{tenenbaum2007,de_beck2013,agundez2014}. 
% Atomic P was also detected in a external galaxy by Molaro et al. (2011) at z=3 using UV spectrography. 
Among the detected P-bearing molecules, PN and PO are the only ones that have been reported in star-forming regions. For decades PN remained as the only P-bearing species observed in these regions \citep{turner1987,ziurys1987,yamaguchi2011,fontani2016}, while PO has been discovered just recently in the surroundings of both high- and low-mass protostars \citep[with PO/PN abundance ratios of $\sim$1$-$3;][]{rivilla2016,lefloch2016}.

Three routes have been proposed for the formation of PN and PO in star-forming regions: (i) shock-induced  desorption of P-bearing species (e.g. PH$_3$) from dust grains and subsequent gas-phase formation \citep{aota2012,lefloch2016}; (ii) high-temperature gas-phase chemistry after the thermal desorption of PH$_3$ from ices (\citealt{charnley94}); and (iii) gas-phase formation of PN and PO during the cold collapse phase and subsequent thermal desorption (at temperatures $\geq$35 K) by protostellar heating \citep{rivilla2016}. Due to the limited number of observations available, and the limited range of physical conditions of the observed regions with detected P-bearing molecules, the formation routes for PN and PO are strongly debated. 
%While \citet{aota2012} and \citet{lefloch2016} claimed a shock-induced enhancement towards the L1157 outflow, this interpretation was challenged by \citet{fontani2016}
%, and recently by Mininni et al. (submitted), 
%based on their PN detections in massive cold and quiescent cores. 

In this Letter, we present observations of PN and PO towards seven regions spread across the Central Molecular Zone (CMZ) in the Galactic Center (GC). These sources are excellent laboratories to test the chemistry of P-bearing molecules since they show different physical properties (high kinetic temperatures, low dust temperatures and moderate densities) and chemistries dominated by either UV photons, cosmic-rays (CR), X-rays or shock waves. 
%To better understand the origin of P-bearing species, we compare PN abundances with that of the well-known shock tracer $^{29}$SiO.  
%Our study shows that the abundance of PN is well correlated with the abundance of $^{29}$SiO, suggesting that P-bearing molecules are sputtered from dust grains due to shocks. The non-detection of PN towards UV-photon/X-ray/cosmic-ray illuminated clouds indicates that this molecule is dissociated by intense radiation.

% Figure of the spatial location of the sources of the sample
%\begin{figure}
%\includegraphics[width=\columnwidth]{figure-Psample.eps}
   %\caption{Positions of the sample of GC sources studied, overplotted on a Spitzer-IRAC 4 image (8 $\mu$m) in color-scale. The different type of sources are shown with different symbols: HMCs (dots), PDRs (stars) and QCs (squares).}
       % \label{fig-sample}
%\end{figure}

\begin{figure*}
\hspace{-0.7cm}
\includegraphics[height=4.525cm, width=14.25cm]{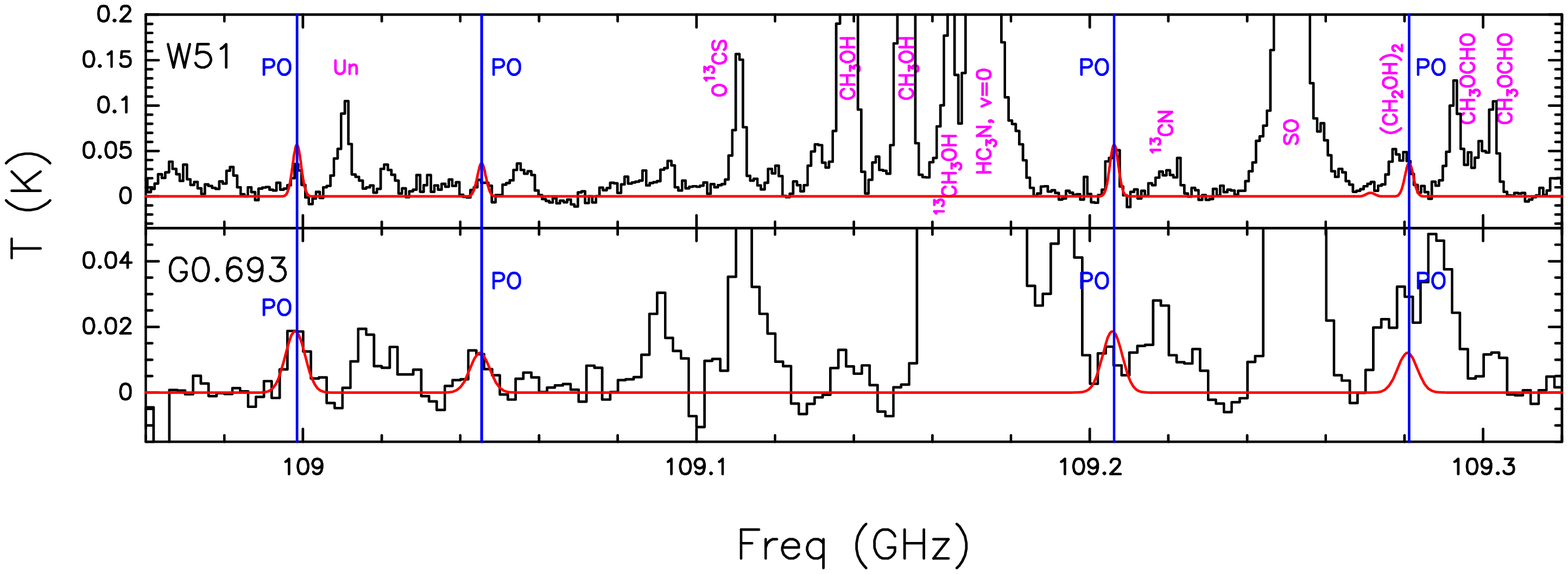}
 \caption{PO detection towards G+0.693-0.03 (lower panel) compared with the detection towards the hot molecular core W51 e1/e2 from \citet{rivilla2016} (upper panel). The PO quadruplet is shown with vertical blue lines. Other molecular species are labeled in the upper panel. The MADCUBA fitted LTE synthetic spectrum of PO in both sources is shown with red lines. }
    \label{fig-PO}
\end{figure*}

%\vspace{-4mm}

\section{The sample}

The CMZ \citep[][]{morris1996} of the Galactic Center harbors the most chemically rich regions in the Galaxy \citep{martin-pintado2001}. The physical conditions in the CMZ are very different from those found in the Galactic disk with high gas temperatures ($\geq$60-100 K), low dust temperatures ($\leq$20 K) and low H$_2$ gas densities \citep[$\sim$10$^4$ cm$^{-3}$;][]{rodriguez-fernandez2000,guesten2004,ginsburg2016}. Since dust is too cold for the evaporation of ices, it has been proposed that the observed chemical richness is due to grain sputtering in widespread, low-velocity shocks (\citealt{martin-pintado97}). In addition, depending on their location, the clouds in the CMZ may be exposed to intense UV radiation (photon-dominated regions or PDRs), X-rays and/or cosmic rays (CR), which strongly affect the chemistry of these clouds \citep{martin2008}. Our selected sample includes two different types of sources (see Table \ref{table-sample}): 

%%%%%%%%%%%%%%%%%%%%%%%%%%%%%%%%%%
%%% GMCs - shock-dominated
%%%%%%%%%%%%%%%%%%%%%%%%%%%%%%%%%%
{\it (i) Shock-dominated regions}: G+0.693$-$0.03, S+0.24+0.01, M$-$0.02$-$0.07 and the molecular rich envelopes of SgrB2 N and M. The first two sources do not show evidence of on-going star formation, while the last three are associated with massive proto-clusters \citep{de_vicente2000,belloche2013,sanchez-monge2017}. In all these regions, SiO is largely enhanced due to shocks \citep{martin2008}.
%and thus their chemistry is likely shock-dominated \citep{martin2008};

%%%%%%%%%%%%%%%%%%%%%%%%%%%%%%%%%%
% GMCs - Radiation dominated
%%%%%%%%%%%%%%%%%%%%%%%%%%%%%%%%%%
{\it (ii) Radiation-dominated regions}: SgrA$^*$ [offset ($-$30$\arcsec$,$-$30$\arcsec$)] and M+0.02$-$0.02.  The first is located at the inner edge of the circumnuclear disk (CND), 1.5 pc away from the black hole. Its molecular gas
%, which present two main velocity components at +20 km s$^{-1}$ and $-$70 km s$^{-1}$ \citep{martin2008,martin2012}, 
is strongly affected by the UV/X-ray radiation (\citealt{amo-baladron11,harada2014}). The CR ionization rate in this region is enhanced by several orders of magnitude \citep{goto2008}, causing a similar effect on the chemistry as that produced by UV photons \citep{harada2015}. 
%It presents two main velocity components: the ambient cloud at +20 km s$^{-1}$ with narrower linewidths, and a second component at -70 km s$^{-1}$ with larger linewidths, even more strongly affected by UV radiation.
M+0.02$-$0.02 is a very compact cloud whose chemistry is
%thought to be affected by supernovae shocks. 
thought to be affected by UV radiation and enhanced CR ionization rate \citep{martin2008}, and also by X-rays (\citealt{ponti2010}). 
%The X-ray ionization rate ($\sim$1$\times$10$^{-14}$ s^-1) is extremely high at the edges of the clouds in the CND (\citealt{harada2013}). 
%M+0.02$-$0.02 cloud might be affected by X-ray emission as it has happened to other Sgr A clouds (Ponti et al. 2010), which are being illuminated by a X-ray flare of Sgr A* fading about 100 years ago.
%\\

%SgrA$^*$ (-30$\arcsec$,-30$\arcsec$)La componente de +20 km/s es la nube ambiente (nube de +20 km/s) con linea estrecha que esta afectada en esta posición por la radiación UV. La componente a -70 km/s mucho mas anchas proviene de CND y esta fuertemente afectada por la radiación UV. Los mapas de HNCO, SiO y CS de Arancha lo muestran claramente. 

%\vskip-3mm
\section{Observations}

The observations were conducted in multiple sessions between 2003 and 2009.
%- October 2005 
%and July 2009 - 
%September 2011 with the IRAM-30m telescope at Pico Veleta, Spain. Table \ref{table-transitions} reports the species and transitions covered within these observations and used in this study.
The observed positions for all sources are listed in Table \ref{table-sample}. In the 2003$-$2005 sessions, the SIS C and D receivers covered the 2 mm window (128$-$176$\,$GHz), whilst between 2009 and 2011 the broad-band EMIR receivers (\citealt{carter2012}) were used at 3mm (80$-$116$\,$GHz). The spectral resolution were in the range 6.8$\,$-9.3 km s$^{-1}$which was high enough to resolve the lines profiles of $\sim$ 20 km s$^{-1}$. The molecular transitions studied in this work are shown in Table \ref{table-transitions}.
%Typical system temperatures ranged between 120-180$\,$K although they could increase up to 390$\,$K at the higher frequencies close to the band edge. 
%The telescope beam sizes were in the range 13$''$ to 31$''$. 
%, estimated using the expression: $\theta$ $_{\rm b}$[arcsec]=2460/$\nu$(GHz). 
%All the spectral line intensities were measured in units of antenna temperature, T$^*_A$. 
Since it is well known that the molecular emission towards these sources is extended over the beam (\citealt{requena-torres2006,martin2008}), in our analysis we have used the line intensities measured in units of antenna temperature, T$^*_A$, as we did in \citet{martin2008} and \citet{requena-torres2008}.  
%T$_{\rm mb}$, using the beam efficiencies reported in the IRAM webpages \footnote{http://www.iram.es/IRAMES/mainWiki/Iram30mEfficiencies}. 
To complete the sample, we also used the publicly available 3 mm spectral survey obtained towards SgrB2 N and M by \citet{belloche2013}.

%  Observations runs  (for G069)
%Aug/Sep 2003
%May/July/Dec 2004
%May/Sep/Oct 2005
%July 2009
%Oct 2010
%Sep 2011

\begin{table*}
\centering
\caption{Sample and results.}
        \tabcolsep 1.5pt
	\begin{tabular}{c c c c c c c c c c c  c c c c} 
    \hline
    \hline
		Source & RA (J2000) & DEC (J2000) & type$^{(a)}$ & Vel. & \multicolumn{2}{c}{FWHM (km s$^{-1}$)} & & \multicolumn{4}{c}{N ($\times$10$^{12}$ cm$^{-2}$)} & [PO/PN] & \multicolumn{2}{c}{T$_{\rm ex}$(K)}\\ \cline{6-7}  \cline{9-12}  \cline{14-15} 
     	 & (h m s) & ($^{\circ}$ $^{\prime}$ $^{\prime\prime}$) & & (km s$^{-1}$) & PN & $^{29}$SiO & &  N(PN) & N(PO) & N($^{29}$SiO)  & N(C$^{34}$S)$^{(b)}$ & &  PN & $^{29}$SiO\\ 	
		\hline
         \hline
		G+0.693$-$0.03     & 17 47 21.86 & $-$28 22 43.00 & Shock & 69 & 24.0 & 24.2 & & 5.6$\pm$0.3  & 8$\pm$3 & 15.5$\pm$0.7 &  50$\pm$7 & 1.5$\pm$0.4 & $\leq$6.3  &  6.8$\pm$0.4$^{(c)}$ \\
        \hline
		S+0.24+0.01   & 17 46 09.86 & $-$28 43 42.39 & Shock & 34 & 12.0 & 21.8 & & 4.8$\pm$0.5 & < 13 & 11.7$\pm$0.2 &  35$\pm$2 & < 2.7 & $\leq$4.2 & 6.3$\pm$0.8$^{(c)}$\\
        \hline
        M$-$0.02$-$0.07  & 17 45 50.64 & $-$28 59 08.81 & Shock & 47 & 19.0 & 27.0 & & 4.7$\pm$0.8  & < 15 & 33.9$\pm$0.7  & 87$\pm$5 & < 3.1  & 5.0$^{(d)}$ & 4.7$\pm$0.5$^{(c)}$\\
        \hline
		SgrB2 N   & 17 47 20.39 & $-$28 22 19.25 & Shock &   64 &  6.7 & 6.6 & & 6$\pm$1  & < 19 &  67.6$\pm$0.8    &  98$\pm$14 & < 1.8  & 5.0$^{(d)}$ & 5.0$^{(d)}$\\
                  &              &             &     &  82  & 9.0 & 9.0 & & 11$\pm$1  & < 19 &  51.3$\pm$0.6     & 26$\pm$8 & < 3.2  & 5.0$^{(d)}$ & 5.0$^{(d)}$\\ 
        \hline
		SgrB2 M   & 17 47 20.41 & $-$28 23 07.25 & Shock & 60 & 6.7 & 6.5 & &  9.3$\pm$0.7  & < 63 &  63$\pm$10   & 129$\pm$22 &  < 6.8 & 5.0$^{(d)}$ & 5.0$^{(d)}$\\	 
                  &             &              &     & 66  & 11.4 & 10.1 & &  18.2$\pm$0.9 & < 63 & 178$\pm$22    & 58$\pm$13 & < 3.5  & 5.0$^{(d)}$& 5.0$^{(d)}$\\
         \hline         
		M+0.02$-$0.02 & 17 45 42.71 & $-$28 55 50.98 & Rad. & 93 & - & 32 & &  < 0.7   & < 11  & 4.9$\pm$0.8 &  25$\pm$6 & -   &5.0$^{(d)}$ & 6.0$\pm$0.9$^{(c)}$\\
 
      &  &  &  & -5.0 & - & 32 & &  < 0.7    & < 11 & 3.8$\pm$0.4 &  34$\pm$15 & -   & 5.0$^{(d)}$ & 6.0$^{(d)}$\\
        \hline   
        
SgrA$^*$(-30$^{\prime\prime}$,-30$^{\prime\prime}$) & 17 45 37.74 & $-$29 00 58.18 & Rad. & 20 & - & - & & < 0.7   & < 18  & < 2.1 &  54$\pm$15 & -  & 5.0$^{(d)}$ & 5.0$^{(d)}$\\  
  &  &  & Rad. &  -70 & - & - &  & < 0.7    & < 18  & < 2.1 &  - $^{(e)}$ & -  & 5.0$^{(d)}$ & 5.0$^{(d)}$\\

		\hline\hline
	\end{tabular}
\label{table-sample}

$^{(a)}$ Shock: Clouds with shock-dominated chemistry; Rad.: Clouds with radiation-dominated chemistry. 
$^{(b)}$  Values extracted from \citet{martin2008} (see their Table 4) for all sources except for SgrB2 N and M (see text).  
$^{(c)}$ The MADCUBA-AUTOFIT was performed using simultaneously the $^{29}$SiO (2$-$1), (3$-$2) and (4$-$3) transitions.
$^{(d)}$ Value fixed. 
$^{e}$ Not derived in \citet{martin2008}.

\end{table*}

% Table

\begin{table}
	\centering
	\caption{PN, PO, $^{29}$SiO and C$^{34}$S transitions used in this work.}
        \tabcolsep 3.0pt
\begin{tabular}{c c c c c} 
		\hline
		Molecule & Transition & Frequency (GHz) & E$_{\rm up}$ (K) & A$_{\rm ul}$ (s$^{-1}$) \\
		\hline
		PN & 2$-$1 & 93.97977 & 6.8  & 2.9$\times$10$^{-5}$ \\
		PN & 3$-$2 & 140.96769 & 13.5 & 1.1$\times$10$^{-4}$ \\
		PO &   F=3$-$2, l=e $^{(a)}$ & 108.99845 &  8.4 & 2.1$\times$10$^{-5}$ \\
		PO &   F=2$-$1, l=e $^{(a)}$ & 109.04540 &  8.4 & 1.9$\times$10$^{-5}$ \\
		PO &  F=3$-$2, l=f  $^{(a)}$   & 109.20620 &   8.4 & 2.1$\times$10$^{-5}$ \\
		PO &  F=2$-$1, l=f  $^{(a)}$ & 109.28119 &  8.4 & 1.9$\times$10$^{-5}$ \\		
        $^{29}$SiO & 2$-$1  & 85.75920  &  6.2 & 2.8$\times$10$^{-5}$ \\	
        $^{29}$SiO & 3$-$2  & 128.63671 &  12.3 & 1.0$\times$10$^{-4}$ \\	
        $^{29}$SiO & 4$-$3  & 171.51229 & 20.6  & 2.5$\times$10$^{-4}$ \\	
        C$^{34}$S  & 2$-$1  & 96.41295 &  6.3 & 1.6$\times$10$^{-5}$ \\	
        C$^{34}$S  & 3$-$2 & 144.61710 &  11.8 & 5.8$\times$10$^{-5}$ \\	
        C$^{34}$S  & 5$-$4 &  241.01609 &  27.0 & 2.8$\times$10$^{-4}$ \\      
%        HNCO & 4$_{0,4}$-3$_{0,3}$ & 87.925237 & 10.5 \\	
		\hline
	\end{tabular}
\label{table-transitions}	

$(a)$ J=5/2$\rightarrow$3/2, $\Omega$=1/2 quadruplet.
%{\bf Shall we include the Aul coefficients?}
%$(b)$ For SgrB2 N we have used the CS(2-1) transition from Belloche et al. (2013) data (see text).
\end{table}

\section{Analysis and Results}

%Table \ref{table-transitions} shows the PN (2$\rightarrow$1) and (3$\rightarrow$2) transitions covered within our survey. 
PN (2$-$1) was detected towards all the shock-dominated clouds in our sample (see Figure \ref{fig-line-profiles} and Table \ref{table-sample}). In the case of SgrB2 N and M, PN (2$-$1) was detected in absorption against the bright continuum emission of the hot molecular cores and compact HII regions, indicating its presence within the low-density envelope of SgrB2. PN was not detected towards the radiation-dominated sources: SgrA$^*$ ($-$30$\arcsec$,$-$30$\arcsec$) and M+0.02$-$0.02. In both sources we have searched at the two velocity components found by \citet{martin2008} in other molecular tracers (see Table \ref{table-sample}). PN (3$-$2) was not detected towards any source.

To derive the column densities of the molecular species across our sample, we have used MADCUBA\footnote{Madrid Data Cube Analysis (MADCUBA) is a software developed in the Center of Astrobiology (Madrid) to visualize and analyze single spectra and datacubes \citep{rivilla2017-1,rivilla2017}.}, which produces synthetic spectra assuming Local Thermodynamic Equilibrium (LTE) conditions.
%and takes the line opacity of each transition into account. %It considers five different parameters to model the expected LTE line profiles: column density of the molecule ({\it N}), excitation temperature ($T_{\rm ex}$), linewidth ($\Delta$v), peak velocity ({\it v}) and source angular diameter ($\theta_{\rm s}$). The MADCUBA AUTOFIT procedure compares the LTE model spectrum with the observed spectrum, by resolving the radiative transfer equation and provides the best non-linear least-squared fit using the Levenberg-Marquardt algorithm. 
%Since the emission is extended and it fills the beam of the IRAM 30$\,$m telescope, no beam dilution was applied to fit the emission lines. 
%For the sources where PN (2$\rightarrow$1) is seen in emission, we considered that the emission is extended, as seen for many other molecular species (Requena-Torres et al. 2006, 2008; Mart\'in et al. 2008), and hence we assume that it fills the telescope beam (i.e. no beam dilution was applied). 
First, we fixed manually the FWHM and the velocity of the lines to values that reproduce well the observed lines (see Table \ref{table-sample}). Then, we used the MADCUBA$-$AUTOFIT tool, which compares the synthetic LTE spectra with the observed spectra and provides the best non-linear least-squared fit using the Levenberg-Marquardt algorithm.
%{\footnote{MADCUBA-AUTOFIT compares synthetic LTE spectra with the observed spectra and provides the best non-linear least-squared fit using the Levenberg-Marquardt algorithm. The parameter errors are derived from the diagonal elements of covariance matrix, the inverse of the Hessian Matrix, and the final chi square of the fit.}} 
We run AUTOFIT to fit simultaneously the PN (2$-$1) detections and the (3$-$2) non-detections.  
The algorithm converged for G+0.693$-$0.03 and S+0.24+0.01, and found low temperatures of 6.3 K and 4.2 K, respectively. 
%We fixed T$_{\rm ex}$ to these values and then rerun AUTOFIT. 
We note that these excitation temperatures should be considered as upper limits since higher values of T$_{\rm ex}$ would excite the PN(3$-$2) transition, which was not detected. 
%Since PN (3$\rightarrow$2) was not detected toward any source, the derived excitation temperatures (T$_{\rm ex}$) of PN should be considered as upper limits.

For M$-$0.02-0.07 the AUTOFIT algorithm did not converge, and thus $T_{\rm ex}$ was fixed to 5 K. 
The derived values for the PN column densities derived with AUTOFIT are shown in Table \ref{table-sample}.
% En las detecciones, lo que hecho es hacer AUTOFIT con la 2-1 y con la 3-2, que me ha dado una Tex. Despues esa la he fijado y he hecho AUTOFIT solo en la 2-1.
The low T$_{\rm ex}$ found are similar to those observed for other species 
%with high critical densities
in these clouds \citep[$\sim$5$-$15 K;][]{requena-torres2008,martin2008}, and to those recently found for PN in several massive cores in the Galactic disk (Mininni et al., subm.). The most likely explanation for the $T_{\rm ex}$ found, much lower than the kinetic temperatures, is that PN is subthermally excited due to the relatively low H$_2$ densities ($\sim$10$^4$ cm$^{-3}$; \citealt{gusten2004}) compared with the critical densities of PN (>10$^5$ cm$^{-3}$, \citealt{tobola07}).

%derived from the collision rate coefficient calculated by \citealt{tobola07}). 

% Fit of the absorption lines
The PN absorption lines towards SgrB2 N and M were fitted assuming that the PN gas is located in a foreground layer with T$_{\rm ex}$=5 K, and considering that the SgrB2 N and M hot cores are background blackbody emitters with sizes of 2$\arcsec$ and gas temperatures of 150 K \citep{belloche2013,sanchez-monge2017}. The PN absorption shows two different velocity components (Figure \ref{fig-line-profiles}) that have been fitted simultaneously. 
%Since no PN (3$\rightarrow$2) emission was detected toward SgrB2N and M, 
%The T$_{\rm ex}$ has been fixed to 5 K, similar to the other sources. 

PN was not detected towards the radiation-dominated regions, SgrA$^*$ ($-$30$^{\prime\prime}$,$-$30$^{\prime\prime}$) and M+0.02-0.02, and then we derived the 3$\sigma$ upper limits for the column density.
%of the two different velocity components present in these sources (Table \ref{table-sample}), assuming $T_{\rm ex}$= 5 K.

%PO
We have also searched for the PO quadruplet at $\sim$109$\,$GHz (Table \ref{table-transitions}). PO has only been detected towards G+0.693$-$0.03, known to be rich in O-bearing molecules \citep{requena-torres2006,requena-torres2008}. Figure \ref{fig-PO} shows the PO detection compared to that reported by \citet{rivilla2016} towards the hot core W51 e1/e2. The PO lines were fitted assuming T$_{\rm ex}$=6.3 K, that inferred from PN. The derived [PO/PN] ratio is $\sim$1.5, similar to that measured in star-forming regions \citep{rivilla2016,lefloch2016}. For the other sources (with PO non-detections), we computed the upper limits of the PO column density considering the same T$_{\rm ex}$ assumed for PN. In all cases, the [PO/PN] ratio is < 7 (Table \ref{table-sample}). 
%For SgrB2N and M, no PO lines were clearly detected, so we present the $3\sigma$ upper limits. 
%The results are shown in Table \ref{table-sample}.

% Fit of SiO
To test whether the chemistry of P-bearing molecules is indeed correlated with the presence of shocks, we have also analyzed the emission of the shock tracer SiO. Since the main isotopologue is optically thick across the GC, we have used the optically thinner isotopologue $^{29}$SiO. By simultaneously fitting the (2$-$1), (3$-$2) and (4$-$3) transitions of this molecule, we have found $T_{\rm ex}$ in the range 4.7$-$6.8 K for G+0.693$-$0.03, S+0.24+0.01, M$-$0.02$-$0.07 and M+0.02$-$0.02.
%(in the latter case, the two velocity components are detected; Table \ref{table-sample}). 
In contrast, this species is not detected in none of the velocity components of SgrA$^*$ ($-$30$^{\prime\prime}$,$-$30$^{\prime\prime}$) and we have thus computed the upper limits. For SgrB2 N and M, the $^{29}$SiO (2$-$1) emission is seen in absorption (Figure \ref{fig-line-profiles}) and, as for PN, we have considered two velocity components and a fixed $T_{\rm ex}$=5 K to compute the column densities (Table \ref{table-sample}).

\begin{figure}
\centering
\includegraphics[height=6.5cm, width=6.5cm]{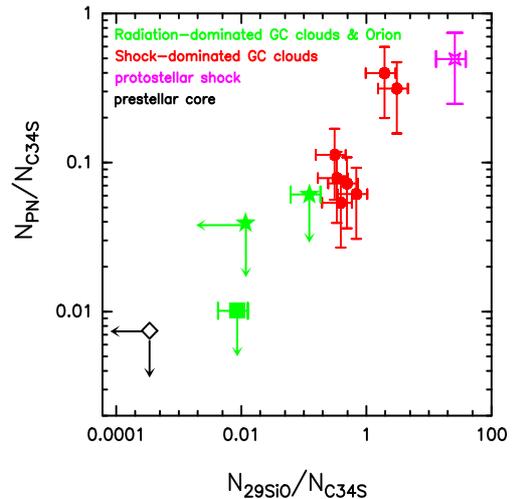}
 \caption{Column density ratios of PN and $^{29}$SiO with respect to C$^{34}$S. The different type of sources are Shock-dominated GC clouds (red dots) and Radiation-dominated GC regions (green stars). The values obtained towards the Orion Bar (a prototypical PDR, green square; Cuadrado, priv. comm.), the L1157-B1 shock (magenta open star) and the L1544 pre-stellar core (black open diamond; from the dataset from \citet{jimenez-serra2016} have also been added. Arrows indicate 3$\sigma$ upper limits. The error bars correspond to conservative uncertainties of $\pm$ half of the molecular ratio value.}
    \label{fig-diagram}
\end{figure}

% Diagram figure
In Figure \ref{fig-diagram}, we plot the column density ratios PN/C$^{34}$S versus those obtained for $^{29}$SiO/C$^{34}$S. The optically-thin C$^{34}$S isotopologue is used here because it is a good proxy of the H$_2$ column density, as shown by \citet{requena-torres2006} in a survey of GC clouds. CS is also a good reference to compute relative molecular ratios because its chemistry is nearly independent of the physical properties of the sources: i) unlike other species (such as CH$_3$OH or HNCO) it is barely enhanced in shocked-gas (\citealt{requena-torres2006}); and ii) it is photoresistant, so it is expected to survive in PDRs (\citealt{requena-torres2006,martin2008}).

For G+0.693$-$0.03, S+0.24+0.01, M$-$0.02$-$0.07, SgrA$^*$ ($-$30$^{\prime\prime}$,$-$30$^{\prime\prime}$) and M+0.02$-$0.02, we have used the C$^{34}$S column densities obtained from C$^{34}$S (3$-$2) and C$^{34}$S (5$-$4) by \citet{martin2008}. For SgrB2 N and M, \citet{martin2008} did not consider the two velocity components used in our analysis 
%of PN and $^{29}$SiO
, and therefore, we have re-done the analysis. For SgrB2 N we have used C$^{34}$S (2$-$1).
%which appears in absorption in the same way as PN and $^{29}$SiO, and that clearly shows two velocity components. 
For SgrB2 M, the C$^{34}$S (2$-$1) line presents strong emission blended with the absorption feature and therefore, we used the C$^{34}$S (3$-$2) transition instead. We used the $T_{\rm ex}$ from \citet{martin2008}, to be consistent with the other sources.
For completeness, in the case of sources with no detection of PN or $^{29}$SiO, we have plotted in Figure \ref{fig-diagram} the values of the upper limits.

% DISCUSSION
Figure \ref{fig-diagram} reveals that there is a positive trend between PN and $^{29}$SiO abundances. 
Although our upper limits toward the PDR-like sources do not allow us to confirm a correlation between PN and $^{29}$SiO, these two species seem to show a similar behaviour: the PN detections correspond to the $^{29}$SiO-richer sources, while PN is not present toward those sources with weaker $^{29}$SiO. 
This strongly suggests that PN is enhanced by shocks. 
This conclusion is in good agreement with the results found by Mininni et al. (subm.) in some massive dense cores of the Galactic disk, and with the shock modeling results from \citet{lefloch2016}, which proposed that PN is formed in gas phase after the shock-induced desorption of PH$_3$.
Since PN is not detected in any GC radiation-dominated cloud%(M+0.02-0.02 and SgrA$^*$ (-30$^{\prime\prime}$,-30 $^{\prime\prime}$))
, this molecule is likely dissociated by the intense UV/X-ray/CR radiation, similarly to HNCO and CH$_3$OH (\citealt{martin2008}). 
For completeness, in Figure \ref{fig-diagram} we also add the PN/C$^{34}$S and $^{29}$SiO/C$^{34}$S column density ratios measured towards the shocked region L1157-B1 \citep{bachiller1997,lefloch2016,podio2017}, the Orion Bar (a prototypical PDR; \citealt{cuadrado2015}, and Cuadrado, priv. comm.), and L1544 \citep[a pre-stellar core without any sign of star formation activity;][]{ward-thompson1999}. In the latter two cases the $^{29}$SiO column density was calculated from SiO assuming $^{28}$Si/$^{29}$Si=19.6 \citep{wilson1999}. For L1544 the molecular column densities were calculated using $T_{\rm ex}$=5 K.
%\footnote{Values obtained from the dataset of \citet{jimenez-serra2016} for the central position of the core. Molecular column densities were calculated using T$_{\rm ex}$=5 K and $^{28}$Si/$^{29}$Si=19.6 \citep{wilson1999}.}. 
The 
%PN/C$^{34}$S and $^{29}$SiO/C$^{34}$S 
column density ratios derived for these regions nicely follow the trend observed for the Galactic Center clouds: PN is %clearly detected and 
well correlated with $^{29}$SiO in the L1157-B1 shock while it remains undetected in the PDR. Since PN is not detected in the L1544 pre-stellar core either 
%(with an upper limit for the abundance of $\leq$4.6$\times$10$^{-13}$)
, this implies that P, like Si, is heavily depleted in molecular dark clouds, in agreement with the shock modelling results from \citet{lefloch2016}, and with observations in molecular dark clouds (\citealt{turner1990}). 
\section{Conclusions}

We searched for P-bearing molecules PN and PO towards seven clouds located in the Galactic Center, known to present different types of chemistry. PN is detected towards five of the seven sources, and % which indicates phosphorus is widespread in the Galactic Center.   
PO is detected only towards one of the sources, G+0.693$-$0.03, which is thought to be the richest source of O-bearing molecules in the Galactic Center. The derived PO/PN abundance ratio is 1.5, similar to values previously found in star-forming regions. The regions where P-bearing species have been detected are clouds thought to be affected by shock waves, and rich in the well-known shock tracer $^{29}$SiO.
The two sources where no P-bearing molecules were detected are regions exposed to intense radiation, and exhibit lower abundances of $^{29}$SiO. %PN shows a good correlation with the shock tracer $^{29}$SiO, which suggests that 
We thus conclude that P-bearing species are formed in the gas phase after the shock-induced sputtering of the grain mantles, and that they
%We have studied the chemistry of P by using different chemical models mimicking the  physical conditions of the different sources: hot molecular cores, quiescent clouds dominated by shocks and PDRs. 
%We also conclude that P-bearing molecules 
are efficiently destroyed by the high cosmic-rays/X-rays/UV-photon radiation expected in some regions of the Galactic Center.

\section*{Acknowledgements}

We thank S. Cuadrado and J. Goicoechea for sharing their data on the Orion bar. V.M.R. is funded by the H2020 programme under the MSC grant agreement No 664931. I.J.-S. and D.Q. acknowledge the financial support received from the STFC through an ER Fellowship and Grant (ST/L004801 and ST/M004139). J.M.-P. acknowledges support by the MINECO grants ESP2013-47809-C03-01 and ESP2015-65597-C4-1. S. Zeng acknowledges support through a Principal's studentship funded by QMUL. DR acknowledge support of the Collaborative Research Council 956, subproject A5, funded by the DFG.
%by the Frederick Perren Fund of the University of London and 

%%%%%%%%%%%%%%%%%%%%%%%%%%%%%%%%%%%%%%%%%%%%%%%%%%



\bibliographystyle{mnras}
\bibliography{Bib} % if your bibtex file is called example.bib

\bsp	% typesetting comment
\label{lastpage}
\end{document}